\documentclass[a4paper,10pt]{article}
\usepackage[utf8x]{inputenc}

\title{Near-extremal black holes}
\author{Bhramar Chatterjee\footnote{bhramar.chatterjee@saha.ac.in}, Amit Ghosh\footnote{amit.ghosh@saha.ac.in}\\ 
\it Saha Institute of Nuclear Physics,\\ \it 1/AF Bidhannagar, Kolkata-700 064, India}

\begin{document}

\maketitle

\begin{abstract}
We present a new formulation of deriving Hawking temperature for near-extremal black holes using distributions. In this paper the near-extremal Reissner-Nordstr\"{o}m and Kerr black holes are discussed. It is shown that the extremal solution as a limit of non-extremal metric is well-defined. The pure extremal case is also discussed separately.
\end{abstract}

\section{Introduction}
Classically a black hole horizon is a one-way membrane that absorbs everything. Since the discovery of Hawking effect \cite{1}, it became clear that quantum
effects cause the black hole to emit thermal radiation. The original derivation of Hawking radiation did not involve a full quantum theory of gravity. 
Only the free fields propagating in a curved spacetime were quantized while the background spacetime remained classical. Several other methods of deriving
Hawking radiation exist in the literature. Recently a semiclassical method of deriving Hawking radiation through tunnelling \cite{2}-\cite{16},\cite{33} has been very 
popular. The tunnelling method involves calculating the imaginary part of the action for the (classically forbidden) process of s-wave emission across the 
horizon, which in turn is related to the Boltzmann factor for emission at the Hawking temperature. Using the WKB approximation the tunnelling probability
for the classically forbidden trajectory of the s-wave coming from inside to outside the horizon is given by:
\begin{equation}\label{semiclassical}
 \Gamma = e^{-2Im S}.
\end{equation}
where $S$ is the classical action of the trajectory to leading order. This is equal to the Boltzmann factor $ e^{-{\beta}E}$ where $\beta$ is the inverse Hawking
temperature. Equating the Boltzmann distribution with (\ref{semiclassical}) implies,
\begin{equation}
 T_{H} = \frac{E}{2Im S}.
\end{equation}
Usually the results are obtained considering a scalar field in the black hole background, but one can also consider fermionic field \cite{17,18,19}. Though the calculations are a bit complicated it is essentially the same.

Like all other formulations the tunnelling picture has its merits and demerits. On the positive sides, this method has been successfully applied to different 
types of horizons including anti-de Sitter(AdS)\cite{24}, de Sitter(dS)\cite{20,21,22,23}, BTZ\cite{25,26},higher dimensional black holes and some exotic spacetimes\cite{27,29,30,31} apart from the more conventional types of black holes,
in each case producing the corresponding Hawking temperature correctly.  Also, as the derivation involves
only the local geometry, the tunnelling method can be applied to any local horizon, in particular to cosmological and weakly isolated horizons \cite{28,29}.
It has also been applied to past horizons and white holes, in which case a clear notion of temperature emerges in complete analogy to black holes \cite{13}. 

On the other hand, the tunnelling formulation gives a less detailed picture of the radiation process since it is mainly related to a semiclassical emission rate.
In this paper we propose a new formulation of the Hawking effect which does not require the WKB approximation to compute the emission rate. Instead we will 
directly construct the modes from the field equation and calculate the emission rate. 
We shall discuss the Reissner-Nordstr\"{o}m(RN) and the Kerr black holes. The idea is to construct single particle states only outside 
the horizon and somehow continue these states to `inside' keeping in mind that the horizon or the surface from where particle escapes to infinity is not a null horizon, 
but something like Hayward's timelike trapping horizon \cite{10,37}. Since this is essentially a horizon crossing phenomenon, a good set of 
coordinates is required which is regular across the horizon, and we will use Kruskal coordinates. Near the horizon the metric is flat and we shall 
construct the field modes using this metric only. In that sense this is also a local calculation of Hawking effect since we need not be bothered about the 
global structure of the spacetime. The outgoing modes have a logarithmic singularity at the horizon but we have to keep in mind that the field modes are 
essentially distributions and it is shown that the distributions are quite well behaved at the horizon. This approach has been previously suggested by Damour and
Ruffini \cite{34}. After constructing the modes, we have calculated 
the probability current coming out of the horizon using the standard field theoretic formula. The conditional probability that a particle emits when it is 
incident on the surface from the other side is then equated to the Boltzmann factor
\begin{equation}
 P_{(emission|incident)} = \frac{P_{(emission{\cap}incident)}}{P_{(incident)}} = e^{-{\beta}E}
\end{equation}
which gives the Hawking temperature associated with the horizon.

The main reason for choosing Kerr and/or RN spacetime is that both exhibit extremal limits. Both metrics have two horizons: the outer $(r = r_+)$ and the inner
$(r = r_-)$ having different properties. They do not even have the same temperature, the temperature of the inner horizon is higher than the outer horizon. The
extremal limit is achieved as $ r_+ \rightarrow r_-$. In this paper we have calculated the probability flux coming out of both the horizon separately and also
what happens in the extremal case. It is found that if one considers the extremal solution as a limiting case of the non-extremal one, then the results agree
with the standard picture, i.e, the temperature of an extremal black hole is zero. On the other hand if one takes an extremal metric from the beginning, then the 
findings are to be interpreted carefully.

This paper is organized as follows: in section 2 we shall discuss Reissner-Nordstr\"{o}m solution, constructing the Kruskal coordinated for both the horizons and
then computing the scalar modes and the probability current for both the cases. Kerr black hole will be considered in section 3, emphasizing again on the 
different sets of Kruskal coordinates for the outer and the inner horizons, the field modes using distributions and finally the probability flux coming out of 
both the horizons. We will show how the extremal solution as a limiting case of the non-extremal one produces the standard results at the end of section 3. 
In section 4 we shall consider an extremal metric and using the aforementioned formulation calculate the probability flux across the 
horizon. 

\section{The Reissner-Nordstr\"{o}m black hole}
\subsection{The metric and Kruskal coordinates}

The Reissner-Nordstr\"{o}m metric is 
\begin{eqnarray}
 ds^2 &=& -f(r)dt^2 + \frac{dr^2}{f(r)} + r^2d\Omega^2.\\
f(r) &=& \left(1-\frac{2M}{r} + \frac{Q^2}{r^2}\right).
\end{eqnarray}
Here, $M$ is the mass and $Q$ is the electric charge of the black hole. The horizons are situated at, $ f(r) = 0$, i.e,
\begin{equation}
 r_{\pm} = M \pm \sqrt{M^2 - Q^2}.
\end{equation}
The outer horizon $r_{+}$ is the event horizon and the inner horizon $r_{-}$ is the apparent horizon for the Reissner-Nordstr\"{o}m spacetime. The extremal limit
is obtained from $ Q \rightarrow M $, or $r_{+} \rightarrow r_{-}$.

The coordinates $t,r$ are singular at the outer horizon $\left(r = r_{+}\right)$ and one can introduce Kruskal-like coordinates to extend the metric across
this surface. However, these coordinates fail to be regular at the inner horizon and so another coordinate system is needed to extend the metric beyond the
inner horizon. So the coordinates are specific to a given horizon and even two coordinate patches fail to cover the entire Reissner-Nordstr\"{o}m manifold. 
Let us consider the two horizons separately.

\subsubsection{The outer horizon}
As, $ r \to r_{+} $,
\begin{equation}
 f(r) \approx 2\kappa_{+}\left(r - r_{+}\right).
\end{equation}
where,\begin{equation}
      \kappa_{+} \equiv\frac{1}{2}{f^{\prime}(r_{+})} = \frac{r_{+} - r_{-}}{2r_{+}^2}. 
      \end{equation}
 is the surface gravity at the outer horizon.
It follows that near $r = r_{+}$, 
\begin{equation}
 r_{*} \equiv \int \frac{dr}{f(r)} \cong \frac{1}{2\kappa_{+}}\ln|\kappa_{+}\left(r - r_{+}\right)|.
\end{equation}
Introducing the null coordinates $u = t - r_{*}$ and $ v = t + r_{*}$, the surface $r = r_{+}$ appears at $ v - u = -\infty$ and we define the Kruskal-like 
coordinates $U_{+}$ and $V_{+}$ by,
\begin{equation}
 U_{+} = {\mp}e^{-\kappa_{+}u},\quad V_{+} = e^{\kappa_{+}v}.
\end{equation}
Here the upper sign refers to $ r > r_{+}$ and the lower sign refers to $ r < r_{+}$.
The future outer horizon is defined as $ U_{+} = 0, V_{+} > 0$.
The metric is regular at the outer horizon as seen from the near horizon form,
\begin{equation}
 ds^2 \simeq -\frac{2}{\kappa_{+}^2}dU_{+}dV_{+} + r_{+}^2d\Omega^2.
\end{equation}
 But $r_{*} \to \infty$ at the inner horizon which is located at $v - u = \infty$ or $U_{+}V_{+} = \infty$ and the Kruskal coordinates are 
singular there. This coordinate patch can be used for $r_{1} < r < \infty$ , where, $r_{1} > r_{-}$. Thus, we need another set of Kruskal coordinates to extend
the spacetime beyond $ r = r_{-}$.

\subsubsection{The inner horizon}
The new set of Kruskal coordinates for the inner horizon can be constructed in a similar manner. As $ r \rightarrow r_{-}$, the function $f(r)$ becomes,
\begin{equation}
 f(r) \approx -2\kappa_{-}\left(r - r_{-}\right).
\end{equation}
where,\begin{equation}
      \kappa_{-} \equiv\frac{1}{2}|{f^{\prime}(r_{-})}| = \frac{r_{+} - r_{-}}{2r_{-}^2}. 
      \end{equation}
Near $r = r_{-}$, 
\begin{equation}
 r_{*} \equiv \int \frac{dr}{f(r)} \cong -\frac{1}{2\kappa_{-}}\ln|\kappa_{-}\left(r - r_{-}\right)|.
\end{equation}
With $ u = t-r_* $  and $ v = t+r_* $, the surface $r = r_{-}$  appears at $ v - u = +\infty$ and we define the new Kruskal coordinates by,
\begin{equation}
 U_{-} = {\mp}e^{\kappa_{-}u} ,\quad V_{-} = -e^{-\kappa_{-}v}.
\end{equation}
Here, the upper sign refers to $r > r_{-}$ and the lower sign refers to $r < r_{-}$. The future inner horizon is defined as
$U_{-} = 0 , V_{-} < 0$.

Then $ f \simeq -2U_{-}V_{-} $ and the metric becomes
\begin{equation}
  ds^2 \simeq -\frac{2}{\kappa_{-}^2}dU_{-}dV_{-} + r_{-}^2d\Omega^2.
\end{equation}
which is regular at $ r = r_{-}$.

\subsection{Scalar modes and the probability current}
A scalar field $\Phi$ satisfies the covariant Klein-Gordon equation and in this background the modes can be separated as,
\begin{equation}
 \Phi_{{\omega}{l}{m}} = \frac{1}{\sqrt{4{\pi}{\omega}}}\frac{\Phi_{\omega}(r_{*},t)}{r}Y_{lm}\left(\theta,\phi \right).
\end{equation}
We shall consider only the positive frequency $(\omega > 0)$ modes which satisfies
\begin{equation}
 i\partial_{t}\Phi_{\omega} = \omega\Phi_{\omega}.
\end{equation}
In Kruskal coordinates (for the outer horizon), the Killing vector $\partial_{t}$ becomes,
\begin{equation}
\partial_{t} = - \kappa_{+}U_{+}\partial_{U_{+}} + \kappa_{+}V_{+}\partial_{V_{+}}.
\end{equation}
So, the $U_{+}$ and $V_{+}$ modes are decoupled,
\begin{equation}
\Phi_{\omega} = \left[f_\omega\left(U_{+}\right) + g_\omega\left(V_{+}\right)\right].
\end{equation}
And the solutions are,
\begin{eqnarray}
 f_\omega\left(U_{+}\right) &=& N_{\omega}|U_{+}|^{\frac{i\omega}{\kappa_{+}}}.\\
g_\omega\left(V_{+}\right) &=& N_{\omega}(V_+)^{-\frac{i\omega}{\kappa_{+}}}.
\end{eqnarray}
where $N_{\omega}$ is the normalization constant.
The $V_{+}$ modes are ingoing in the outer horizon and is well behaved across the horizon.

The $U_{+}$ modes are outgoing and are not well behaved close to the horizon because they oscillate infinitely rapidly. However, to calculate the emission 
probability we shall need these modes only.

The probability current is positive definite for positive frequency modes and associated with the $U_{+}$ modes it is
\begin{equation}
 j^{out} = -i\left[-\kappa_{+}U_{+}\left(\partial_{U_{+}}\overline{\Phi_{\omega}}\right)\Phi_{\omega}
 + \kappa_{+}U_{+}\overline{\Phi_{\omega}}\left(\partial_{U_{+}}\Phi_{\omega}\right)\right].
\end{equation}
The $U_{+}$ modes are defined inside and outside the horizon, but as it approaches the horizon at $U_+ = 0$ the modes pick up a logarithmic singularity and is 
not differentiable. So $j^{out}$ cannot be calculated naively. But actually, these modes are distribution valued as mentioned earlier and not to be interpreted 
as ordinary functions. As distributions, they are well defined and infinitely differentiable at the horizon. The distributions are of the form \cite{35},
\begin{equation}
 f_{\omega} = \lim_{\epsilon \to 0}N_{\omega}|U_{+} + i\epsilon|^{\frac{i\omega}{\kappa_{+}}}
= \left\{ \begin{array}{ll}
           N_{\omega}\left(U_{+}\right)^{\frac{i\omega}{\kappa_{+}}} & \mbox{ for $U_{+} > 0$,} \\  N_{\omega}|U_{+}|^{\frac{i\omega}{\kappa_{+}}}
e^{-\frac{\pi\omega}{\kappa_{+}}} & \mbox{ for $U_{+} < 0$. } 
            \end{array}
\right.
\end{equation}
and the complex conjugate distribution is,
\begin{equation}
  \overline{f_{\omega}} = \lim_{\epsilon \to 0}N_{\omega}^*\,|U_{+} - i\epsilon|^{-\frac{i\omega}{\kappa_{+}}}
= \left\{ \begin{array}{ll}
       N_{\omega}^*\left(U_{+}\right)^{-\frac{i\omega}{\kappa_{+}}} & \mbox{ for $U_{+} > 0$,} \\ N_{\omega}^*\,|U_{+}|^{-\frac{i\omega}
{\kappa_{+}}}
e^{-\frac{\pi\omega}{\kappa_{+}}} & \mbox{ for $U_{+} < 0$. } 
            \end{array}
\right.
\end{equation}
These distributions are uniquely associated with the $U_+$-modes if we impose the additional condition that these are well behaved for large frequencies, 
$\omega\to \infty$. The derivatives of the distributions are also uniquely determined,
\begin{eqnarray}
 \partial_{U_{+}}f_\omega &=& N_{\omega}\left(\frac{i\omega}{\kappa_+}\right)\lim_{\epsilon \to 0}|U_{+} + i\epsilon|^{\frac{i\omega}{\kappa_{+}}-1}.\\
\partial_{U_{+}}\overline{f_\omega} &=& N_{\omega}^{*}\left(-\frac{i\omega}{\kappa_+}\right)\lim_{\epsilon \to 0}|U_{+} - i\epsilon|^{-\frac{i\omega}
{\kappa_{+}}-1}.
\end{eqnarray}
So, the probability current associated with the outgoing modes is,
\begin{equation}
 j^{out} = \omega U_+ |N_\omega|^2\lim_{\epsilon \to 0}\left[\frac{1}{U_+ -i\epsilon}+\frac{1}{U_+ + i\epsilon}\right]\left(U_+ 
-i\epsilon\right)^{-\frac{i\omega}{\kappa_+}}\left(U_+ + i\epsilon\right)^{\frac{i\omega}{\kappa_+}}.
\end{equation}
Now, $U_+\left(U_+ \pm i\epsilon\right)^{-1}$ gives the identity distribution, because $\left(U_+ \pm i\epsilon\right)^{-1} = PV(1/U_+) \mp i\pi\delta(U_+)$
and $U_+\delta(U_+) = 0$. Finally,
\begin{equation}
\lim_{\epsilon\to 0}\left(U_+\mp i\epsilon\right)^{\mp\frac{i\omega}{\kappa_+}}=\lim_{\epsilon\to 0} e^{\mp\frac{i\omega}{\kappa_+}\ln\left(U_+\mp\,i\epsilon
\right)}=e^{\mp\frac{i\omega}{\kappa_+}(\ln|U_+|\mp\,i\pi\theta(-U_+))}.
\end{equation}
As a result, we get the outgoing probability current,
\begin{equation}
 j^{out} = \left\{ \begin{array}{ll}
           |N_{\omega}|^{2} & \mbox{ for $U_{+} > 0$,} \\  |N_{\omega}|^{2}e^{-\frac{2\pi\omega}{\kappa_{+}}} & \mbox{ for $U_{+} < 0$. } 
            \end{array}
\right.
\end{equation}
In a similar manner, using the Kruskal coordinates on the inner horizon one can calculate the outgoing probability current inside and outside of the inner 
horizon. In this case,
 \begin{equation}
 j^{out} = \left\{ \begin{array}{ll}
           |\tilde{N_{\omega}}|^{2} & \mbox{ for $U_{-} > 0$,} \\  |\tilde{N_{\omega}}|^{2}e^{-\frac{2\pi\omega}{\kappa_{-}}} & \mbox{ for $U_{-} < 0$. } 
            \end{array}
\right.
\end{equation}

\section{Kerr black hole}
\subsection{The metric and Kruskal coordinates}
The Kerr metric is given by,
\begin{equation}
 ds^{2} = -\left(1-\frac{2Mr}{{\rho}^2}\right)dt^2 - \frac{4Mar{\sin^2{\theta}}}{\rho^2}dtd\phi + \frac{\Sigma}{\rho^2}{{\sin^2{\theta}}}d\phi^2 + 
\frac{\rho^2}{\Delta}dr^2 + {\rho^2}d\theta^2.
\end{equation}
where,
\begin{eqnarray}
 \rho^2 &=& r^2 + a^2{\cos}^2\theta. \\
\Delta &=& r^2 + a^2 - 2Mr.\\
\Sigma &=& \left(r^2 + a^2\right)^2 - a^2\Delta{\sin}^2\theta. 
\end{eqnarray}
The horizons are situated at $\Delta = 0 $, i.e at,
\begin{equation}
 r_{\pm} = M \pm \sqrt{M^2 - a^2}.
\end{equation}
Construction of Kruskal coordinates for the Kerr spacetime is very similar to the Reissner-Nordstr\"{o}m spacetime though a bit more complicated.
Let us define the $ u,v $ coordinates for the Kerr spacetime by using the following transformations,
\begin{eqnarray}
 r_{*} &=& \int\frac{r^2 + a^2}{\Delta}dr.\\
\tilde\phi &=& \phi - \frac{at}{2Mr_{+}}.
\end{eqnarray}
And as usual,
\begin{eqnarray}
u &=& t - r_{*}.\\
v &=& t + r_{*}.
\end{eqnarray}
Let us now define the Kruskal like coordinates for the Kerr spacetime,
\begin{equation}
U_{+} = {\mp}e^{-\kappa_{+}u},\quad V_{+} = e^{\kappa_{+}v}.
\end{equation}
where
\begin{eqnarray}
U_{+}V_{+} &=& {\mp}e^{2\kappa_{+}r}\left(r - r_{+}\right)\left(r - r_{-}\right)^{-\frac{\kappa_{+}}{\kappa_{-}}}.\\
\kappa_{+} &=& \frac{1}{2}\left(\frac{r_{+} - r_{-}}{r_{+}^2 +a^2}\right). \\
\kappa_{-} &=& \frac{1}{2}\left(\frac{r_{+} - r_{-}}{r_{-}^2 +a^2}\right) .
\end{eqnarray}
$\kappa_{+}$ is the surface gravity at the outer horizon. Here again, the upper sign is for $r > r_{+}$, and the lower sign is for $r < r_{+}$. The future outer
horizon is defined as $ U_{+} = 0, V_{+} > 0$.

The metric near the horizon$\left(r \to r_{+}, U_{+} \to 0\right)$ in the Kruskal coordinates takes the form
\begin{eqnarray*}
 ds^2 &=& {\mp}4\rho_{+}^2\left(r_{+} - r_{-}\right)^{\frac{\kappa_{+}}{\kappa_{-}}-1}e^{-2\kappa_{+}r_{+}}dU_{+}dV_{+} + \rho_{+}^2d\theta^2\\
&+& 4a^2{\sin^2\theta}\left(r_{+} -r_{-}\right)^{2\frac{\kappa_{+}}{\kappa_{-}}-2}\left(\frac{r_{+}^2}{\rho_{+}^2} + \frac{r_{+}^2-a^2}{r_{+}^2+a^2}\right)
e^{-4\kappa_{+}r_{+}}V_{+}^2dU_{+}^2\\
&\pm& 2a{\sin^2\theta}\left(r_{+} -r_{-}\right)^{\frac{\kappa_{+}}{\kappa_{-}}}\left(1 + \frac{r_{+}}{\kappa_{+}\rho_{+}^2}\right)
e^{-2\kappa_{+}r_{+}}V_{+}dU_{+}d\tilde\phi \\
&+& \frac{\left(r_{+}^2 + a^2\right)^2}{\rho_{+}^2}{\sin^2\theta}d\tilde\phi^2.
\end{eqnarray*}
So, the coordinate singularity is removed and the metric is regular at the horizon in the $U_{+},V_{+}$ coordinates.

Just as for the RN spacetime, the coordinates $\left(U_{+},V_{+}\right)$ are singular at the inner horizon $(r = r_{-})$, and another coordinate patch is 
required to extend the Kerr metric beyond this horizon.

For the inner horizon we define
\begin{eqnarray}
  r_{*} &=& \int\frac{r^2 + a^2}{\Delta}dr.\\
\tilde\psi &=& \phi - \frac{at}{2Mr_{-}}.
\end{eqnarray}
With $u = t-r_*$ and $v = t+r_*$, the appropriate choice for Kruskal coordinates are,
\begin{equation}
 U_{-} = {\mp}e^{\kappa_{-}u},	 V_{-} = -e^{-\kappa_{-}v} .
\end{equation}
The upper sign refers to $r>r_{-}$ and the lower sign refers to $r<r_{-}$. Near the horizon $\left(r \rightarrow r_{-}, U_{-} \rightarrow 0\right)$ the metric
becomes
\begin{eqnarray*}
 ds^2 &=& {\mp}4\rho_{-}^2\left(r_{+} - r_{-}\right)^{\frac{\kappa_{-}}{\kappa_{+}}-1}e^{2\kappa_{-}r_{-}}dU_{-}dV_{-} + \rho_{-}^2d\theta^2\\
&+& 4a^2{\sin^2\theta}\left(r_{+} -r_{-}\right)^{2\frac{\kappa_{-}}{\kappa_{+}}-2}\left(\frac{r_{-}^2}{\rho_{-}^2} + \frac{r_{-}^2-a^2}{r_{-}^2+a^2}\right)
e^{4\kappa_{-}r_{-}}V_{-}^2dU_{-}^2\\
&\pm& 2a{\sin^2\theta}\left(r_{+} -r_{-}\right)^{\frac{\kappa_{-}}{\kappa_{+}}}\left(\frac{r_{+}}{\kappa_{+}\rho_{+}^2} -1\right)
e^{2\kappa_{-}r_{-}}V_{-}dU_{-}d\tilde\psi \\
&+& \frac{\left(r_{-}^2 + a^2\right)^2}{\rho_{-}^2}{\sin^2\theta}d\tilde\psi^2.
\end{eqnarray*} 
The metric is regular at $r = r_{-}$, and the spacetime can be extended beyond this horizon.

\subsection{Scalar modes and the probability current}
For the Kerr metric the scalar modes can be separated as \cite{36}
\begin{equation}
 \Psi_{{\omega}m} = \frac{1}{\sqrt{4\pi\omega}}\frac{\Psi_{\omega}\left(r_*,t\right)}{r}e^{im\phi}\Theta\left(\theta\right).
\end{equation}
The Killing vector in this case is $\left(\partial_t + \Omega_{H}\partial_{\phi}\right)$ and considering only the positive frequency solutions as before we get
\begin{equation}
  i\left(\partial_{t} + \Omega_{H}\partial_{\phi}\right)\Psi_{{\omega}m} = \omega\Psi_{{\omega}m}.
\end{equation}
Here, $ \omega = E - m\Omega_{H}$ where $E = \omega_{\infty}$ is the frequency at infinity.

In Kruskal coordinates the Killing vector becomes
\begin{eqnarray}
\partial_{t} + \Omega_{H}\partial_{\phi}  &=& \partial_u +\partial_v + \frac{\partial\tilde\phi}{\partial t}\partial_{\tilde\phi} + \Omega_H \frac{\partial\tilde\phi}{\partial\phi}\partial_{\tilde\phi}\\
 &=& - \kappa_{+}U_{+}\partial_{U_{+}} + \kappa_{+}V_{+}\partial_{V_{+}} .
\end{eqnarray} 
just the same as the RN spacetime, and hence the $U_{+}$ and the $V_+$ modes are decoupled,
\begin{equation}
 \Psi_\omega = \left[f_{\omega}(U_+) + g_{\omega}(V_+)\right].
\end{equation}
 Concerning ourselves with only the outgoing $U_+$ modes, we get the
solutions 
\begin{equation}
f_{\omega}\left(U_{+}\right) = N_{\omega}|U_{+}|^{\frac{i\omega}{\kappa_{+}}}.
\end{equation}
Again as before, these modes are well behaved both inside and outside the horizon and is not differentiable at the horizon because of the logarithmic singularity.
And we have to resort back to the distributions to calculate the probability current for emission. The distributions have the same form as in the case of RN 
spacetime, and the probability current through the outer horizon is
\begin{equation}
 j^{out} = \left\{ \begin{array}{ll}
           |N_{\omega}|^{2} & \mbox{ for $U_{+} > 0$,} \\  |N_{\omega}|^{2}e^{-\frac{2\pi\omega}{\kappa_{+}}} & \mbox{ for $U_{+} < 0$. } 
            \end{array}
\right.
\end{equation}
For the inner horizon, using the specific Kruskal coordinates, the probability current is the same as the RN inner horizon
 \begin{equation}
 j^{out} = \left\{ \begin{array}{ll}
           |\tilde{N_{\omega}}|^{2} & \mbox{ for $U_{-} > 0$,} \\  |\tilde{N_{\omega}}|^{2}e^{-\frac{2\pi\omega}{\kappa_{-}}} & \mbox{ for $U_{-} < 0$. } 
            \end{array}
\right.
\end{equation}

\subsection{Extreme limit of non-extremal solutions}

A non-extremal spacetime with an outer $(r_+)$ and an inner $(r_-)$ horizon becomes extremal as $r_+\rightarrow r_-$. The two horizons are in equilibrium at two
different temperatures and as a result, the outgoing fluxes are also different. As was shown earlier, both for the Reissner-Nordstr\"{o}m and Kerr metrics, the
temperature of the outer horizon is, $T_{out} = \frac{\hbar\kappa_+}{2\pi}$ and that of the inner horizon is $T_{in} = \frac{\hbar\kappa_-}{2\pi}$. 
Since $k_->k_+$ the inner horizon is in equilibrium at a higher temperature than that outer one. So the outgoing flux from the outer horizon, given by (30) 
and (52) for the R-N and Kerr spacetimes respectively, are less than the incoming flux through the inner horizon given by (31) and (53) respectively. 
The results are consistent with expectations. In the extremal limit as $r_+ \to r_-$, $T_{out} \to 0$ as $\kappa_+ = 0$. So no thermality is observed at the outer horizon as expected. This can be shown more clearly by considering an effective temperature for the outer horizon.

For both RN and Kerr black holes, since the spacetime between the two horizons for $(r_- < r < r_+)$ is vacuum, the fluxes have to match. This implies,
\begin{equation}
|\tilde{N_{\omega}}|^{2}e^{-\frac{2\pi\omega}{\kappa_{-}}} = |N_{\omega}|^{2}.
\end{equation}
 So, the effective flux coming out of the outer horizon is given by,
\begin{equation}
 j^{out} = |N_{\omega}|^{2}e^{-\frac{2\pi\omega}{\kappa_{+}}} = |\tilde{N_{\omega}}|^{2}e^{-2\pi\omega\left(\frac{1}{\kappa_+}+\frac{1}{\kappa_-}\right)}.
\end{equation}
This gives an effective temperature of the outer horizon as the harmonic mean of $\kappa_+$ and $\kappa_-$,
\begin{equation}
\beta_{eff} = \frac{2\pi}{\hbar}\left(\frac{1}{\kappa_+} + \frac{1}{\kappa_-}\right).
\end{equation}
Far from extremality, $M\gg Q$, as a result $r_+ \gg r_-$ and $\kappa_-\gg\kappa_+$. Thus, $T_{eff} \approx T_{out}$.

But in the extremal limit, as $r_- \rightarrow r_+$, $T_{eff} = 0 $.
This shows that the outgoing flux approaches zero not only on the outer horizon but on the inner horizon as well. This limit is consistent with what we expect 
from an extremal solution.

\section{Extremality}
The nature of the extremal metric is quite different from other stationary solutions. Still we can calculate the scalar modes and the outgoing probability 
current through the horizon following the same procedure as that of a stationary metric. Naturally, the mode solutions near the horizon is different and this 
leads to a different type of distributions for the extremal case. We found that though the flux vanishes precisely at the horizon, leading to a zero
temperature for the horizon, it is non-zero both outside and inside of the horizon, which is not physically acceptable. However, no such problems arise if we 
consider the extremal spacetime as a limiting case of a non-extremal one.     

\subsection{Extremal solution}
The extreme Reissner-Nordstr\"{o}m metric is
\begin{equation}
 ds^2 = -\left(1-\frac{M}{r}\right)^2 dt^2 + \frac{dr^2}{\left(1-\frac{M}{r}\right)^2} + r^2 d\Omega^2.
\end{equation}
where $M$ is the mass of the black hole. The horizon is at $r = M$. 
Introducing the null coordinates $u = t - r_*$, $v= t + r_*$ ,with
\begin{equation}
 r_* \equiv \int\frac{dr}{\left(1-\frac{M}{r}\right)^2} = r + 2M\ln|r - M| - \frac{M^2}{r-M}.
\end{equation}
the surface $r = M$ appears at $v-u = -\infty$. The Kruskal coordinates $U,V$ are given by the 
implicit relations,
\begin{eqnarray}
 u &=& -M\cot U.\\
v &=& - M\tan V.
\end{eqnarray}
The future horizon is located at $U = 0, V < \frac{\pi}{2}$.

The Killing vector $\partial_t$ is
\begin{equation}
 \partial_t = \frac{1}{M}\left[\sin^2{U}\partial_U - \cos^2{V}\partial_V\right].
\end{equation}
Again the $U$ and the $V$ modes are decoupled and the positive frequency solutions for a scalar field are found to be
\begin{eqnarray}
\Psi_{\omega}(U) &=& N_{\omega}e^{i\omega M\cot U} .\\
\Psi_{\omega}(V) &=& N_{\omega}e^{i\omega M\tan V}.
\end{eqnarray}
The ingoing $V$-modes are regular across the horizon. Near the horizon the outgoing $U$-modes takes the form
\begin{equation}
 \Psi_{\omega}(U) = N_{\omega}e^{i\omega M\cot U} \simeq N_{\omega}e^{\frac{i\omega M}{U}}.
\end{equation}
The singularity at the horizon can be removed by using distributions. In this case the appropriate distributions are found by taking logarithm of the modes, 
\begin{eqnarray}
 \ln \Psi_{\omega}(U) &=& \ln N_{\omega} + \lim_{\epsilon\rightarrow 0}\frac{i\omega M}{U -i\epsilon}.\\
\ln \overline\Psi_{\omega}(U) &=& \ln N_{\omega}^* - \lim_{\epsilon\rightarrow 0}\frac{i\omega M}{U + i\epsilon}.
\end{eqnarray}
The extremal solution for the Kerr spacetime is obtained by setting $a = M$ in the Kerr metric. The line element becomes,
\begin{equation}
 ds^2 = - \left(1-\frac{2Mr}{{\rho}^2}\right)dt^2 - \frac{4M^2r{\sin^2{\theta}}}{\rho^2}dtd\phi + \frac{\Sigma}{\rho^2}{{\sin^2{\theta}}}d\phi^2 + 
\frac{\rho^2}{\Delta}dr^2 + {\rho^2}d\theta^2.
\end{equation}
with,
\begin{eqnarray}
 \rho^2 &=& r^2 + M^2{\cos}^2\theta. \\
\Delta &=& \left(r-M\right)^2 .\\
\Sigma &=& \left(r^2 + M^2\right)^2 - M^2\Delta{\sin}^2\theta. 
\end{eqnarray}
The horizon is situated at $r = M$. Introducing the null coordinates $u = t - r_*, v = t + r_*$ as before, with
\begin{eqnarray}
 r_* &\equiv& \int\frac{r^2 + M^2}{\left(r-M\right)^2}dr = r + 2M\ln|r - M| - \frac{2M^2}{r-M}.\\
\tilde\phi &=& \phi - \frac{1}{2M}t = \phi - {\Omega}t
\end{eqnarray}
the surface $r = M$ appears at $v-u = -\infty$. The Kruskal coordinates $U,V$ are the same as in the case of extremal Reissner-Nordstr\"{o}m metric,
\begin{eqnarray}
 u &=& -M\cot U.\\
v &=& - M\tan V.
\end{eqnarray}
The future horizon is located at $U = 0, V < \frac{\pi}{2}$.

The Killing vector $\partial_t + \Omega\partial_{\phi}$ is
\begin{eqnarray}
 \partial_t + \Omega\partial_{\phi} &=& \partial_u + \partial_v + \frac{\partial\tilde\phi}{\partial t}\partial_{\tilde\phi} + \Omega\partial_{\tilde\phi}\\
    &=&\frac{1}{M}\left[\sin^2{U}\partial_U - \cos^2{V}\partial_V\right].
\end{eqnarray}

So, the mode solutions for the extremal Kerr metric are similar to the Reissner-Nordstr\"{o}m case. 

To find the outgoing probability flux across the horizon, we have calculated
\begin{eqnarray}
 \frac{j^{out}}{\overline\Psi_{\omega}(U)\Psi_{\omega}(U)} &=& -\frac{iU^2}{M}\partial_U\left[\ln \overline\Psi_{\omega}(U) - \ln \Psi_{\omega}(U)\right]\\
 &=& \omega U^2\left[\frac{1}{(U +i\epsilon)^2} + \frac{1}{(U-i\epsilon)^2}\right].
\end{eqnarray}
Now, $\left(U\pm i\epsilon\right)^{-2} = PV(1/U^2) \pm i\pi\delta^\prime(U)$ and $U^2\delta^\prime(U) = -2U\delta(U) = 0$, so the above quantity is equal to 
$2\omega$.
Finally taking logarithm of both sides again we get the flux
\begin{equation}
 \ln j^{out} = \ln |N_\omega|^2 - 2\pi\omega M\delta(U).
\end{equation}
Clearly, for all $N\neq 0$, $j^{out} = |N_\omega|^2$, namely no thermality is observed. It is as if the horizon is a transparent membrane. To calculate its 
`value' at $U=0$, we regularize the delta-function; in the limit $\epsilon \to 0$
\begin{equation}
 j^{out} = |N_\omega|^2\exp\left(-2M\omega\frac{\epsilon}{U^2 + \epsilon^2}\right).
\end{equation}
Then $j^{out} = |N_\omega|^2\exp\left(-2M\omega/\epsilon\right) $ which approaches zero in the limit $\epsilon\to 0$. So there is a finite discontinuity of 
flux at the horizon and no thermal effect is noted. However, one can argue that this discontinuity is a coordinate artefact. Recall that for an extremal black hole, be it an extremal Reissner-Nordstr\"om or extremal Kerr, the proper radial distance from the horizon to any point however close to the horizon outside or inside, is infinite. Thus, it is impossible for any particle state to cross the horizon when incident from inside or outside. This is the basic reason why an extremal black hole cannot absorb or emit particle states. The coordinates $U_+,V_+$ remove the coordinate singularity at the horizon. But unlike the non-extremal cases, $U_+$ fails to be a proper distance. So a discontinuity in $U_+$ is not a physical discontinuity. The flux gets infinite proper time to become zero at the horizon when it is incident either from inside or outside. When equated to the Boltzmann factor, it implies an infinite $\beta$. This is equivalent to setting a zero temperature for the black hole.

\section{Conclusion and discussions}
We have investigated Hawking radiation for near extremal Reissner-Nordstr\"{o}m and Kerr black holes from a different viewpoint. First, we have developed a method of calculating temperature associated with a horizon based on quantum field theory
in a curved background. We have constructed the field modes (in this case a scalar field) near the horizon and computed the probability current coming out of the horizon using the modes. The problem of horizon crossing is taken care of by the distributional nature of these modes. This can be regarded as an improvement over the conventional tunnelling calculations since we do not need to calculate any semiclassical action using WKB approximation to find the emission probability. These approximations are never precisely established in the conventional analysis apart from the fact that they give the `correct' Hawking temperatures. Our analysis shows that the results are more robust and independent of such approximations. The only approximation we have used is the flatness of the metric near the horizon which is justifiably so in well-defined coordinates.  

Second, we have considered radiation from both the outer and the inner horizon
for these two spacetimes. It was found that the two horizons are in thermal equilibrium at different temperatures, the temperature of the inner horizon is 
higher than that of the outer horizon, as a result the incoming flux through the inner horizon is less than the outgoing flux through the outer horizon.
Matching the current in between the two horizons we have found an effective temperature for the outer horizon which is such that when the spacetime is far from
extremality $(r_+ \gg r_-)$ the effective temperature is the temperature of the outer horizon for massive black holes. On the other hand, in the extremal limit, when $r_+ \to r_-$, the effective temperature is zero (actually, both temperatures vanish in this limit) and the emission probability is also zero. This is consistent with the expectations.

In the case of an extremal metric it was found that the flux is the same both outside and inside the horizon and zero at the horizon. This indicates a zero temperature for the pure extremal black hole, though there is a finite discontinuity of flux which is not observed in the extremal limit of the non-extremal spacetimes. In this sense the extremal spacetime as the limiting case of the non-extremal one is more physically acceptable. 

In the past, extremal black holes have been considered by several authors in the context of tunnelling and contradictory results were obtained. Both in the ``Hamilton-Jacobi'' variant of the tunnelling approach and the ``null geodesic'' approach the real part of the action $S$ is divergent. In the Hamilton-Jacobi approach one gets a vanishing imaginary part \cite{7} (which naively implies a zero $\beta$, hence a divergent Hawking temperature) and in the ``null geodesic'' approach the imaginary part of the action is non-vanishing and the temperature of an extremal black hole is found to be ``quantized in units of the temperature of a Schwarzschild black hole'' \cite{11}. These lead to the conclusions that as if the tunnelling formulation breaks down for extremal cases because it fails to reproduce the expected results. However, in our formulation such complications do not arise because our method does not involve any semi-classical approximation or action. In the distributional sense our result is consistent with the standard viewpoint. 
 
\centerline{\bf Acknowledgement}\vspace{0.5cm}

We acknowledge discussions with Parthasarathi Mitra.


\begin{thebibliography}{99}
%
\bibitem{1} Hawking S W 1975 {\it Commun. Math. Phys.} {\bf 43} 199-220
\bibitem{2} Hartle J B and Hawking S W 1976 {\it Phys. Rev. D}{\bf 13} 2188
\bibitem{3} Parikh M K and Wilczek F 2000 {\it Phys. Rev. Lett.} {\bf 85} 5042–5
\bibitem{4} Srinivasan K and Padmanabhan T 1999 {\it Phys. Rev. D} {\bf 60} 24007
\bibitem{5} Shankaranarayanan S, Srinivasan K and Padmanabhan T 2001 {\it Mod. Phys. Lett. A} {\bf 16} 571–8
\bibitem{6} Shankaranarayanan S, Padmanabhan T and Srinivasan K 2002 {\it Class. Quantum Grav.} {\bf 19} 2671–88
\bibitem{7} Angheben M, Nadalini M, Vanzo L and Zerbini S 2005 {\it J. High Energy Phys.} {\bf 05(2005)} 014 
\bibitem{8} Nadalini M, Vanzo L and Zerbini S 2006 {\it J. Phys. A: Math. Gen.} {\bf 39} 6601 
\bibitem{9} Di Criscienzo R, Nadalini M, Vanzo L, Zerbini S and Zoccatelli G 2007 {\it Phys. Lett. B} {\bf 657} 107–11 
\bibitem{10} Di Criscienzo R, Hayward S, Nadalini M, Vanzo L and Zerbini S 2009 {\it Class. Quantum Grav.} {\bf 26} 062001 
\bibitem{11} Kerner R and Mann R B 2006 {\it Phys. Rev. D} {\bf 73} 104010
\bibitem{12} Mitra P 2007 {\it Phys. Lett. B} {\bf 648} 240-2
\bibitem{13} Chatterjee B, Ghosh A and Mitra P 2008 {\it Phys. Lett. B} {\bf 661} 307–11
\bibitem{14} Banerjee R and Majhi B R 2009 {\it Phys. Lett. B} {\bf 675} 243–5
\bibitem{15} Visser M 2003 {\it Int. J. Mod. Phys. D} {\bf 12} 649
\bibitem{16} Nielsen A B and Visser M 2006 {\it Class. Quantum Grav.} {\bf 23} 4637
\bibitem{17} Kerner R and Mann R B 2008 {\it Class. Quantum Grav.} {\bf 25} 095014, {\it Phys. Lett. B} {\bf 665} 277-283 
\bibitem{18} Di Criscienzo R and Vanzo L 2008 {\it Europhys. Lett.} {\bf 82} 600001 
\bibitem{19} Yale A 2011 {\it Phys. Lett. B} {\bf 697} 398-403 
\bibitem{20} Parikh M K 2002 {\it Phys. Lett. B} {\bf 546} 189 
\bibitem{21} Medved A J M 2002 Phys. Rev. D 66 124009 
\bibitem{22} Shankaranarayanan S 2003 Phys. Rev. D 67 084026;
Wu S Q and Jiang Q Q 2006, arXiv:hep-th/0603082;
Chen D Y, Jiang Q Q and Zu X T 2008, arXiv:0804.0131 [hep-th]
\bibitem{23} Stotyn S, Schleich K and Witt D 2009 Class. Quantum Grav. 26 065010 
\bibitem{24} Hemming S and Keski-Vakkuri E 2001 {\it Phys. Rev. D} {\bf 64} 044006 
\bibitem{25} Medved A J M 2002 {\it Class. Quantum Grav.} {\bf 19} 589–98 
\bibitem{26} Li R and Ren J R 2008 {\it Phys. Lett. B} {\bf 661} 370–2 
\bibitem{27}Banerjee R and Modak S K 2009 {\it J. High Energy Phys.} {\bf 11} 073 
\bibitem{28} Wu X and Gao S 2007 {\it Phys. Rev. D} {\bf 75} 044027 
\bibitem{29} Medved A J M 2002 {\it Phys. Rev. D} {\bf 66} 124009 
\bibitem{30} Kerner R and Mann R B 2007 {\it Phys. Rev. D} {\bf 75} 084022 
\bibitem{31} Ren J, Zhao Z and Gao C 2006 {\it Gen. Rel. Grav.} {\bf 38} 387–92 
%\bibitem{32} de Gill A, Singleton D, Akhmedova V and Pilling T 2010 Am. J. Phys. 78 685–91 
\bibitem{33} L Vanzo, G Acquaviva and R Di Criscienzo 2011 {\it Class. Quantum Grav.}{\bf 28} 183001 
\bibitem{34} Damour T and Ruffini R 1976 {\it Phys. Rev. D} {\bf 14} 332–4
\bibitem{35} Gel'fand I.M and Shilov G.E 1964 {\it Generalized Functions V1} {(Academic Press)}
\bibitem{36} Ford L.H 1975 {\it Phys. Rev. D} {\bf 12} 2963
\bibitem{37} Chatterjee A, Chatterjee B, Ghosh A {\em in preparation}

\end{thebibliography}
\end{document}